\documentclass[%
 aip,
 amsmath,amssymb,
 reprint,%
groupedaddress
]{revtex4-1}

\usepackage{placeins}
\usepackage{graphicx}
\usepackage{dcolumn}
\usepackage{bm}

\usepackage[utf8]{inputenc}
\usepackage[T1]{fontenc}
\usepackage{mathptmx}
\usepackage{etoolbox}
\usepackage{circuitikz}
\usepackage{hyperref}
\usepackage{subcaption}
\usepackage{balance}
\usepackage{tikz}

\def\cavity#1#2{ 
\draw (#1-2cm,#2-2cm) rectangle (#1+2cm,#2+2cm);
\draw (#1-2.5cm,#2-2.5cm) rectangle (#1+2.5cm,#2+2.5cm);
\fill [fill=gray] (#1-0.1cm,#2-2cm) rectangle (#1+0.1cm,#2-1cm);}

\def\crystal#1#2{ 
\fill [color=blue!60] (#1-1cm,#2+1cm) to (#1-1cm,#2-1cm) to (#1+1cm,#2-1cm) to (#1+1cm,#2+1cm);
\draw (#1+0cm,#2+0cm) node [scale=2] {LiF};}

\def\labelledbox#1#2#3#4#5{ 
\filldraw [fill=white] (#1-#3*0.5,#2+#4*0.5) rectangle (#1+#3*0.5,#2-#4*0.5);
\draw (#1,#2) node [scale=1.2] {#5};}

\def\attenuator#1#2#3{ 
\filldraw [fill=white] (#2-0.5,#3-0.5) rectangle (#2+0.5,#3);
\draw (#2,#3-0.25) node [scale=0.7] {#1 dB};}

\def\isolator#1#2{
\filldraw [fill=white] (#1-0.4,#2-1) rectangle (#1+0.4,#2);
\draw [-{Triangle[scale=1.5]}] (#1,#2-0.9) to (#1,#2-0.1);}

\def\amplifier#1#2#3#4{
\filldraw [fill=white,rotate around={#1:(#2,#3)}] (#2-0.4,#3) to (#2+0.4,#3) to (#2,#3+0.8) to (#2-0.4,#3);
\draw [rotate around={#1:(#2,#3)}] (#2,#3+0.3) node {#4};}

\def\supmagnet#1#2{
\filldraw [fill=white] (#1-4.5,#2-2) rectangle (#1-3.5,#2+2);
\filldraw [fill=white] (#1+3.5,#2-2) rectangle (#1+4.5,#2+2);
\draw (#1-4,#2+1.5) circle [radius=0.23];
\draw (#1-4,#2+0.5) circle [radius=0.23];
\draw (#1-4,#2-0.5) circle [radius=0.23];
\draw (#1-4,#2-1.5) circle [radius=0.23];
\draw (#1+3.8,#2+1.7) to (#1+4.2,#2+1.3);
\draw (#1+4.2,#2+1.7) to (#1+3.8,#2+1.3);
\draw (#1+3.8,#2+0.7) to (#1+4.2,#2+0.3);
\draw (#1+4.2,#2+0.7) to (#1+3.8,#2+0.3);
\draw (#1+3.8,#2-0.7) to (#1+4.2,#2-0.3);
\draw (#1+4.2,#2-0.7) to (#1+3.8,#2-0.3);
\draw (#1+3.8,#2-1.7) to (#1+4.2,#2-1.3);
\draw (#1+4.2,#2-1.7) to (#1+3.8,#2-1.3);}

\def\coordsystem#1#2{
\draw[->, thick, >=Stealth] (#1,#2) to (#1 - 0.5547002, #2 - 0.83205029);
\draw [left] (#1 - 0.5547002, #2 - 0.83205029) node [scale=1.5] {$\mathbf{x}$};
\draw[->, thick, >=Stealth] (#1,#2) to (#1,#2 + 1);
\draw [above] (#1,#2 + 1) node [scale=1.5] {$\mathbf{z}$};
\draw[->, thick, >=Stealth] (#1,#2) to (#1 + 1,#2);
\draw [right] (#1 + 1,#2) node [scale=1.5] {$\mathbf{y}$};}

\def\dtensor#1#2#3#4#5#6#7#8{
\draw[->, thick, >=Stealth] (#1,#2) to (#1 - #3*0.5547002, #2 - #3*0.83205029);
\draw [left] (#1 - #3*0.5547002, #2 - #3*0.83205029) node [scale=1.5] {#6};
\draw[->, thick, >=Stealth] (#1,#2) to (#1,#2 + #4*1);
\draw [above] (#1,#2 + #4*1) node [scale=1.5] {#7};
\draw[->, thick, >=Stealth] (#1,#2) to (#1 + #5*1,#2);
\draw [right] (#1 + #5*1,#2) node [scale=1.5] {#8};}

\def\Bfieldup#1#2{
\draw[->, thick, >=Stealth] (#1,#2) to (#1,#2 + 3);
\draw [right] (#1, #2*0.5 + 3*0.5) node [scale=2] {$\mathbf{B}$};
\draw [left] (#1, #2*0.5 + 3*0.5) node [scale=2] {$\mathbf{z}$};}

\makeatletter
\def\@email#1#2{%
 \endgroup
 \patchcmd{\titleblock@produce}
  {\frontmatter@RRAPformat}
  {\frontmatter@RRAPformat{\produce@RRAP{*#1\href{mailto:#2}{#2}}}\frontmatter@RRAPformat}
  {}{}
}%
\makeatother
\begin{document}

\preprint{AIP/123-QED}

\title{Cryogenic Microwave Whispering Gallery Mode Spectroscopy of Paramagnetic Impurities in High-Purity Crystalline LiF}

\author{Steven Samuels}
\thanks{Corresponding author email: 22720743@student.uwa.edu.au}
\author{William Campbell}
\author{Michael E. Tobar}
\author{Maxim Goryachev} %
\affiliation{%
 Quantum Technologies and Dark Matter Lab \\
 The University of Western Australia, 35 Stirling Hwy, Crawley WA 6009
}

\date{\today}

\begin{abstract}
  A low-noise cryogenic microwave spectroscopy experiment was performed on a high-purity lithium fluoride (LiF) crystal. The spectroscopy data revealed avoided level crossing interactions in whispering gallery modes, indicative of electron spin resonance (ESR) coupling with paramagnetic impurities. Analysis of the interaction spectra identified distinct spin systems corresponding to $(S = 3/2, I = 7/2)$, $(S = 1, I = 7/2)$, and $(S = 3/2, I = 0)$. The number of hyperfine splittings observed, together with the natural abundance of ions possessing the appropriate nuclear spin values, suggest that V$^{2+}$ and V$^{3+}$ impurities, exhibiting orthorhombic distortion, are the most likely sources of the narrow interaction features. This interpretation is supported by earlier ESR studies and established manufacturing records for LiF crystal growth. Additionally, a separate set of broader interaction points is consistent with an orthorhombic model involving a $(S = 3/2, I = 0)$ spin system, although the specific impurity responsible for this interaction remains unidentified.
  \end{abstract}

\maketitle


\section{Introduction}

The detection and identification of impurities in crystal dielectrics serve as critical steps in the development of high precision devices and experiments \cite{bowman1988detection, zaripov5039546new, neill2025compensation, subramanian2025czochralski}. A powerful technique for detecting minute populations of impurity ions in low-loss dielectric crystals is cryogenic microwave spectroscopy \cite{farr2013ultrasensitive, hartman2024precision, hosain2018whispering}. Building on earlier ESR spectroscopy studies of LiF, we further explore its impurity structure using cryogenic microwave spectroscopy.

LiF is an ionic compound with applications ranging from optical devices to fundamental physics experiments \cite{basiev1994room, mirov2001alexandrite, ter1995efficient, baldacchini1998colour, gellermann1991color, ellis1991elastic, rossiter1971titanium, hanafee1967effect, li2012charge}. Small concentrations of atomic impurities in crystalline structures can influence macroscopic material properties such as conductivity, luminescence, and elasticity. Therefore, characterizing crystal impurities is crucial. A prominent example is the use of LiF as a bolometer for dark matter particle detection \cite{minowa1993direct}. This detector relies on low energy nuclear recoils caused by the elastic scattering of hypothesized dark matter particles. Impurities in the crystal can reduce the detector’s sensitivity to these recoils, introducing unknown noise into the experiment. 

In optical devices, F$^{-}$ impurity defect sites, known as colour centers in LiF, can serve as pump sources for lasers operating near the infrared spectrum \cite{basiev1994room, mirov2001alexandrite}. The efficiency of the lasing transition will depend on the concentration and type of colour center impurities in the crystal \cite{mirov2001alexandrite, janesko2018time}. 

In other applications, dilute impurities in LiF are essential to the operation of low noise precision microwave devices and experiments for the purposes of quantum measurement, computation, and control \cite{creedon2010high, creedon2011high, locke2000monolithic}. It is well known that qubits with long coherence times can be realised from dilute crystal spin defects such as nitrogen vacancy centers in diamond \cite{probst2013anisotropic, amsuss2011cavity, nguyen2025first}.

Research on impurity concentrations in LiF remains limited. Therefore, characterizing the crystalline structure of typical LiF samples is well justified for various physics applications. In this work, we conduct low noise microwave spectroscopy on pure LiF for the purpose of characterising magnetic material impurities. Such impurities may exist in the crystal lattice at parts per million levels \cite{farr2013ultrasensitive}. 

In Section II, we describe the experimental setup. In Section III, we present the mathematical model for simulating ESR spectra. Section IV presents the agreement between simulation results and experimental data suggesting the presence of a population of charged impurity ions.

\section{Experiment}

The identification of spin impurities is achieved by measuring the frequency and amplitude response of whispering gallary modes (WGM) in a microwave cavity containing the crystal. In this technique, a cylindrical dielectric crystal is placed inside a microwave cavity. The crystal's low dielectric loss and excellent confinement of photonic modes in the bulk allow it to host high quality factor (denoted as Q) WGMs. The polarisation of these WGMs can be categorised into two types: transverse magnetic polarisation (WGH) or transverse electric polarisation (WGE) with respect to the cylindrical z-axis \cite{giordano2023degeneracy}. A constant external magnetic field is then applied to the crystal and is ramped to high field amplitudes. Electron spin resonance (ESR) is the phenomenon in which the energy of the microwave photons of a certain frequency matches the Zeeman splittings due to a spin transition of host impurities. Thus, incoming microwave photons will be absorbed, causing the impurity ion to transition to a higher spin state. Assuming the presence of a small population of impurity ions, the WGMs will undergo frequency shifts and amplitude changes in the form of avoided level crossings \cite{PhysRevLett.110.157001} near magnetic field values where resonance occurs. Thus, such impurity ions can be identified by searching for their corresponding Zeeman lines in the host crystal's microwave response.

The crystal and the cavity are kept at a temperature close to 10 mK through liquid helium dilution to exploit high Q-factors of WGM resonators at cryogenic temperatures \cite{farr2013ultrasensitive, hartman2024precision, hosain2018whispering, bourhill2019low}. Higher Q factors will lead to longer photon lifetimes, increasing SNR of spin resonance transitions. Other factors which can affect the strength of interaction between WGMs and spin impurities are concentration, mode volume and mode polarisation \cite{PhysRevLett.110.157001}.

\begin{figure}
  \begin{center}
      \scalebox{0.48}{\begin{tikzpicture}
          \draw [right] (3.5,5.1) node {MXC};
          \draw [right] (3.5,6.6) node {COLD};
          \draw [right] (3.5,7.6) node {STILL};
          \draw [right] (3.5,9.1) node {4K};
          \draw [right] (3.5,12.1) node {60K};
          \draw [right] (3.5,13.1) node {TOP};
          \draw [right] (0,19) node [scale=1.5] {Field control};
          \draw [left] (-1,17) node [scale=1.5] {Measured $S_{21}$};
          \draw [right] (1,17) node [scale=1.5] {Measurement};
          \draw (1.4,0) circle (1mm);
          \draw (1.5,0) to (2,0);
          \draw [-{>[scale=3]}] (2.5,0) to (3,0) to (3,9.6) to (3,11.6) to (3,14);
          \draw (3,13) to (3,14.5) to (1.5,14.5) to (1.5,15.5);
          \draw (-1.4,0) circle (1mm);
          \draw (-1.5,0) to (-2,0);
          \draw [-{<[scale=3]}] (-2.5,0) to (-3,0) to (-3,14);
          \draw (-3,13) to (-3,14.5) to (-1.5,14.5) to (-1.5,15.5);
          \draw [-{>[scale=3]}] (-3,20) to (-5.5,20) to (-5.5,0) to (-4.5,0);
          \draw [-{>[scale=3]}] (3,20) to (5.5,20) to (5.5,0) to (4.5,0);
          \draw [-{>[scale=3]}] (-1,16.5) to (-1,17.5);
          \draw [-{>[scale=3]}] (1,17.5) to (1,16.5);
          \draw [-{>[scale=3]}] (0,18.5) to (0,19.5);
          \cavity{0}{0}
          \crystal{0}{0}
          \draw [dashed] (-5,-3) rectangle (5,13.5);
          \filldraw [fill=white]  (-0.5,2.5) rectangle (0.5,5);
          \filldraw [fill=white]  (-3.5,5) rectangle (3.5,5.2);
          \filldraw [fill=white] (-3.5,6.5) rectangle (3.5,6.7);
          \filldraw [fill=white] (-3.5,7.5) rectangle (3.5,7.7);
          \filldraw [fill=white] (-3.5,9) rectangle (3.5,9.2);
          \filldraw [fill=white] (-3.5,12) rectangle (3.5,12.2);
          \filldraw [fill=white] (-3.5,13) rectangle (3.5,13.2);
          \attenuator{-20}{-3}{5}
          \attenuator{-10}{-3}{9}
          \isolator{3}{5}
          \amplifier{0}{3}{10.3}{}
          \supmagnet{0}{0}
          \labelledbox{0}{16}{4}{1}{VNA}
          \labelledbox{0}{18}{3}{1}{PC}
          \labelledbox{0}{20}{6}{1}{Magnet Power Supply Controller}
          \fill [fill=blue!10, fill opacity=0.4]
          (-5,13.5) rectangle (5,9.5);
          \fill [fill=blue!30, fill opacity=0.4] (-5,9.5) to (5,9.5) to (5,-3) to (-5,-3);
          \draw [dashed] (-5,9.5) to (5,9.5);
          \fill [fill=blue!40, fill opacity=0.4] (-4,8.2) to (4,8.2) to (4,5.5) to (-4,5.5);
          \draw [dashed] (-4,5.5) to (-4,8.2) to (4,8.2) to (4,5.5);
          \fill [fill=blue!50, fill opacity=0.4] (-3.2,-3) to (-3.2,2.5) to (-4,2.5) to (-4,5.5) to (4,5.5) to (4,2.5) to (3.2,2.5) to (3.2,-3);
          \draw [dashed] (-3.2,-3) to (-3.2,2.5) to (-4,2.5) to (-4,5.5) to (4,5.5) to (4,2.5) to (3.2,2.5) to (3.2,-3);
          \Bfieldup{7}{0}
      \end{tikzpicture}}
  \end{center}
  \caption{Experimental setup for observing magnetic spin interactions in a helium dilution refrigerator. Regions with different temperatures are shaded different saturations of blue with higher saturations representing lower temperatures. Notice that the z-axis of the crystal is aligned with the direction of the magnetic field.}
  \label{fig:LiF_experimental_setup}
\end{figure}
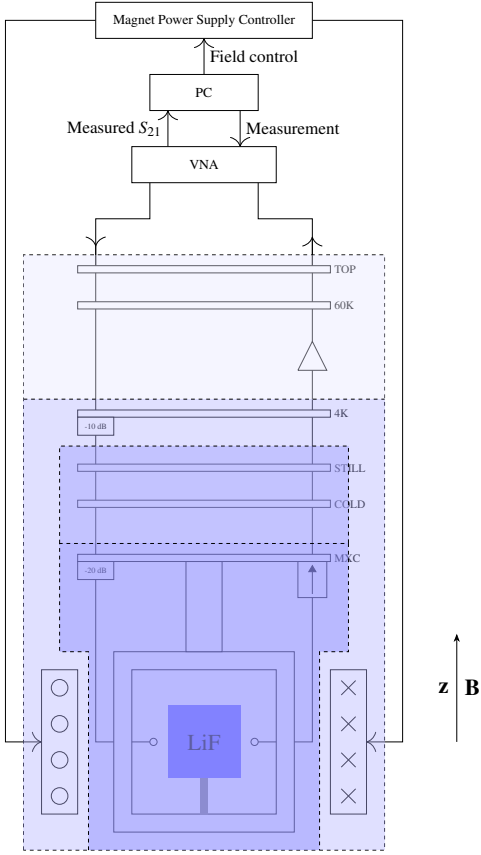

The LiF cylinder was grown epitaxially in the z direction using the Czochralski technique then cut along the (001) plane. It has a diameter of 30 mm and a thickness of 30 mm. In addition, a 3 mm by 15 mm hole is precisely drilled into the bottom of the crystal to accommodate a copper rod from the cavity. Looped conductors are attached on opposite sides of the copper cavity and are responsible for coupling RF signals into and out of the cavity. The position of the probes are adjusted such that reflected power from the cavity due to impedence mismatch is minimized. The cylindrical cavity itself is manufactured from oxygen-free high thermal conductivity copper. This is to attain high quality factor WGMs at low temperatures by removing hot spots from oxygen impurities that increase surface losses.

A vector network analyzer is used to measure the frequency response of the transmitted power, $S_{21}(\omega)$, through the WGMs in the resonator at excitation powers ranging from -90 dBm to -30 dBm. Higher powers are avoided, as they may heat the crystal and introduce nonlinear effects, which are undesirable for ESR measurements. Cold attenuators were connected on the input side of the 4 K plate and the mixing chamber plate in order to prevent the 300 K noise from entering the cavity. The output of the cavity is connected to an isolator then a 4 K low noise cryogenic amplifier. The isolator prevents broadband noise from the amplifier input from propagating back into the microwave cavity through its output port.

The frequency stability of the measurement system is ensured through the use of a hydrogen maser reference clock.

\section{Theoretical Model of the Experiment}

The general Hamiltonian for a spin interacting with an external magnetic field is given by the following \cite{griffiths2019introduction}:

\begin{equation} \label{eq:genspinham}
  \hat{H} = \sum_{i} \mathbf{B}_{i} \cdot \mathbf{\hat{S}}_{i} + \sum_{i,j} \mathbf{J}_{ij} \mathbf{\hat{S}}_{i} \cdot \mathbf{\hat{S}}_{j}
\end{equation}

Here, $\mathbf{B}_{i}$ is the magnetic field at the site of the $i$th spin, $\mathbf{\hat{S}}_{i}$ is the spin operator for the $i$th spin, and $\mathbf{J}_{ij}$ is the exchange interaction between the $i$th and $j$th spins. ESR resonance occurs when the energy difference between two eigenstates of (\ref{eq:genspinham}) matches the energy of a microwave photon from a WGM:

\begin{equation}\label{eq:resonance}
  \hbar f_{WG} = \Delta E = E_{m} - E_{n}
\end{equation}

Here, $f_{WG}$ is the frequency of the WG mode and $\Delta E$ is the energy difference between the $m$th and $n$th eigenstates of (\ref{eq:genspinham}). For specific impurity populations, an orthorhombic spin model can be considered. In this model, it is sufficient to consider only the electron Zeeman interaction, zero-field splitting interaction, and the hyperfine interaction as the super-hyperfine interaction is negligible \cite{chan1971electron}. For arbitrary orientations of orthorhombic distortion, the spin interaction Hamiltonian can be written as follows~\cite{nehrkorn2015simulating}:

\begin{align} \label{eq:spin_ham_arborient}
    \hat{H} &= \mu_{B} g_{e} B_{z} \hat{S}_{z} + h \mathbf{\hat{S}}^{T} \cdot \mathbf{D} \cdot \mathbf{\hat{S}} + h A \mathbf{I} \cdot \mathbf{S}
\end{align}

\begin{figure*}[t]
  \centering
  \includegraphics[width=\textwidth]{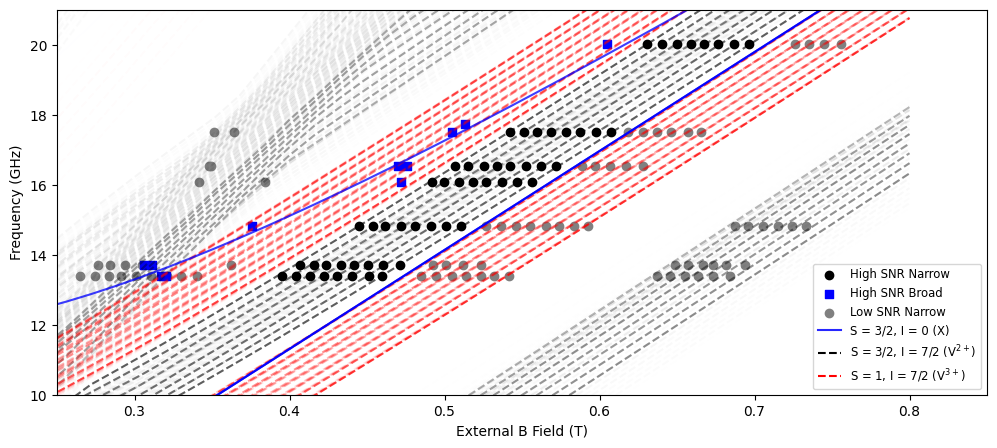}
  \caption{Simulated EPR spectra generated using equation (\ref{eq:spin_ham_arborient}) with parameter values for $g$, $E$, $D$, and $A$ taken from Table \ref{tab:parameter_values}. The spectra are fitted to selected interaction points and are categorized by their spectroscopic signal-to-noise ratio (SNR). Simulations assume different electron and nuclear spin numbers for distinct sets of data points. Each spectrum is assigned either a most likely impurity type or labeled as unknown (denoted by X).}
  \label{fig:spin_transitions_rotations}
\end{figure*}

Here, the first term represents the electron Zeeman interaction, the second term corresponds to the zero-field splitting interaction, and the third term describes the hyperfine interaction. For orthorhombic distortions, $\mathbf{D}$ can be oriented to the principal axis to reduce the zero-field splitting term in Equation~(\ref{eq:spin_ham_arborient}) to the following \cite{hutton1969rotational, konig1976zero}:

\begin{align}\label{eq:principle_axis_D}
  \mathbf{\hat{S}}^{T} \cdot \mathbf{D} \cdot \mathbf{\hat{S}} &= D \left( S_{z}^{2} - \frac{1}{3}S(S+1) \right) + E(S_{x}^{2} - S_{y}^{2})
\end{align}

In EPR spectroscopy, transition probabilities determine the signal intensity. According to time-dependent perturbation theory, the probability of an EPR transition $\left| i \right\rangle \rightarrow \left| f \right\rangle$ is given by \cite{abragam2012electron}:

\begin{align}
  \rho_{i,f} &\propto \left| \bm{B}_{\text{WGM}} \cdot \left< f \left| \hat{\bm{\mu}} \right| i \right> \right|^{2} (p_{i} - p_{f}) \\ \nonumber
  \hat{\bm{\mu}} &= -g\mu_{B}\hat{\bm{S}}
\end{align}

Here, $\bm{B}_{\text{WGM}}$ denotes the magnetic field component of the WGM, and $\hat{\bm{\mu}}$ represents the magnetic dipole operator, while $p_{f}$ and $p_{i}$ denote the populations of the final and initial states, respectively. The external magnetic field defines the quantization axis so only components of the spin operator perpendicular to this axis, specifically $\hat{S}_{x}$ and $\hat{S}_{y}$, contribute to the transition probability. Transitions are therefore considered forbidden when the following condition is satisfied:

\begin{align} \label{eq:probability}
   \left| \left< f \right| \hat{S}_{x} \left| i \right> \right|^{2} + \left| \left< f \right| \hat{S}_{y} \left| i \right> \right|^{2} = 0
\end{align}

This condition can be used to filter out forbidden transitions from the simulated spectra.

\section{Main Results and Discussions}

\begin{figure}[h]
    \centering
    \begin{subfigure}{0.48\textwidth}
      \centering
      \includegraphics[width=\textwidth]{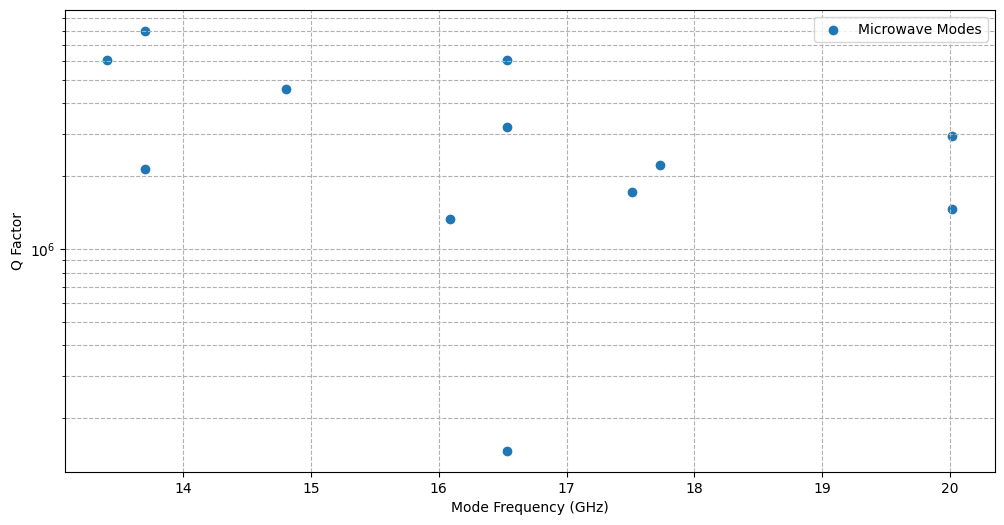}
      \caption{Q factor of WGMs involved in the spectroscopy.}
      \label{fig:example_mode_Q}
    \end{subfigure}
    \hfill
    \begin{subfigure}{0.48\textwidth}
        \centering
        \includegraphics[width=\textwidth]{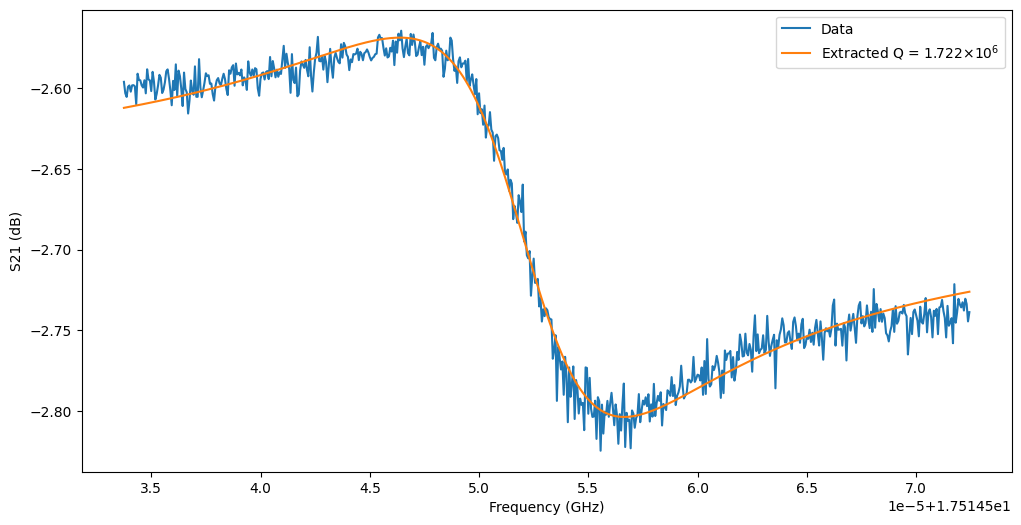}
        \caption{Example trace of one of the WGMs from above.}
        \label{fig:Q_example_trace}
    \end{subfigure}
    \caption{Q factors of WGMs involved in the spectroscopy and an example trace.}
    \label{fig:Q_factors_plots}
\end{figure}

The Q-factors of WGMs used in this spectroscopy study range from 0.1 million to 8 million, as shown in Figure \ref{fig:example_mode_Q}. The Q-values obtained in this experiment are two orders of magnitude lower than those from a similar microwave spectroscopy experiment on sapphire reported in \cite{farr2013ultrasensitive}. As a result, the interactions are weaker and not always easily observed due to poor SNR.

In Figure~\ref{fig:spin_transitions_rotations}, the interaction points corresponding to various WGH modes, identified using the measurement setup described above, are plotted. The black circular data points, which indicate WGM interactions, appear in groups of eight. This characteristic Zeeman structure provides strong evidence for a host impurity system in LiF. The most likely spin system responsible for producing eight interaction points per group involves a nuclear spin of $I = 7/2$. Various oxidation states of vanadium were determined to be the most likely candidates for the observed interactions due to their relatively high natural abundance. All circular data points in Figure~\ref{fig:spin_transitions_rotations} are therefore fitted using either the V$^{2+}$ ($S = 3/2$) or the V$^{3+}$ ($S = 1$) ion impurity models.

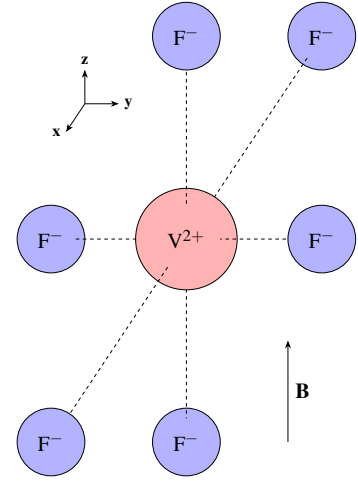
\begin{figure}[t]
  \begin{center}
      \scalebox{0.40}{\begin{tikzpicture}
          \filldraw [fill=blue!30] (4,6) circle (1);
          \draw (4,6) node [scale=2] {F$^{-}$};
          \filldraw [fill=blue!30] (0,-6) circle (1);
          \draw (0,-6) node [scale=2] {F$^{-}$};
          \filldraw [fill=blue!30] (-4,0) circle (1);
          \draw (-4,0) node [scale=2] {F$^{-}$};
          \draw [dashed] (0,0) to (3.61170986, 5.41756479);
          \draw [dashed] (0,0) to (0,-5.3);
          \draw [dashed] (0,0) to (-3.3,0);
          \filldraw [fill=red!30] (0,0) circle (1.5);
          \draw (0,0) node [scale=2] {V$^{2+}$};
          \draw [dashed] (-0.5547002, -0.83205029) to (-4,-6);
          \draw [dashed] (0,6) to (0,1);
          \draw [dashed] (4,0) to (1,0);
          \filldraw [fill=blue!30] (0,6) circle (1);
          \draw (0,6) node [scale=2] {F$^{-}$};
          \filldraw [fill=blue!30] (4,0) circle (1);
          \draw (4,0) node [scale=2] {F$^{-}$};
          \filldraw [fill=blue!30] (-4,-6) circle (1);
          \draw (-4,-6) node [scale=2] {F$^{-}$};
          \draw[->, thick, >=Stealth] (3,-6) to (3,-3);
          \draw [right] (3,-4.5) node [scale=2] {$\mathbf{B}$};
          \coordsystem{-2.5}{2.5}
          \dtensor{4}{2.5}{1}{1}{1}{$\mathbf{D}_{zz}$}{$\mathbf{D}_{yy}$}{$\mathbf{D}_{xx}$}
      \end{tikzpicture}}
  \end{center}
  \caption{Orthorhombic distortion of a V$^{2+}$ ion in the LiF crystal lattice. The surrounding fluoride ions are slightly displaced along the crystal axes, resulting in three distinct distances to the central V$^{2+}$ ion, as opposed to the uniform distances found in a cubic field. The orientation of the zero-field tensor in its principal axis frame is also shown (right) for all impurities listed in Table~\ref{tab:parameter_values}. Each axis of the tensor is labeled as $\mathbf{D}_{xx}$, $\mathbf{D}_{yy}$, or $\mathbf{D}_{zz}$.}
  \label{fig:orthorhombic_distortion}
\end{figure}

During Czochralski crystal growth, trace impurities such as vanadium and cobalt—both possessing a nuclear spin of $I = 7/2$—can be unintentionally incorporated into the sample at low concentrations. A well-documented case of this occurs in the growth of GaAs, where vanadium originating from a boron nitride crucible was found to contaminate the crystal \cite{brandt1985identification}. The most general form of vanadium crystal field distortion in LiF is orthorhombic \cite{chan1970electron}. Therefore, the Hamiltonian in equation~(\ref{eq:spin_ham_arborient}) was used to simulate the spin system. This Hamiltonian was diagonalized to obtain the energy levels and corresponding EPR spectra, which are represented by the dashed lines in Figure~\ref{fig:spin_transitions_rotations}. Forbidden transitions can be filtered out using equation~(\ref{eq:probability}). The parameters $g$, $E$, $D$, and $A$ were treated as free variables to fit the simulated spectra to the interaction points observed in the experimental data. The resulting $\mathbf{D}$ tensor was then rotated into its principal axis frame (see Figure~\ref{fig:orthorhombic_distortion} for the orientation of the $\mathbf{D}$ tensor for all impurities relative to the laboratory frame). The best-fit values, along with those reported in \cite{chan1971electron}, are listed in Table~\ref{tab:parameter_values}. 

Note that some of the parameter values obtained in this ESR experiment differ slightly from those reported in~\cite{chan1971electron}. These discrepancies may arise from fundamental differences in the defect environment and crystal geometry relative to the applied magnetic field in ESR spectroscopy. For example, different charge compensation mechanisms—such as the substitution of V$^{2+}$ for Li$^{+}$, which introduces a net $+1$ charge—may be compensated by F$^{-}$ vacancies, Li$^{+}$ interstitials, or nearby impurity ions such as OH$^{-}$ or O$^{2-}$~\cite{watkins1959electron}. These mechanisms can distort the crystal lattice in various ways, modifying the local symmetry and thereby affecting the $g$-tensor and zero-field splitting parameters, and/or altering the covalency, which in turn influences the hyperfine coupling constant.

\begin{table}[h]
  \caption{\label{tab:parameter_values} Parameters ($g$, $D$, $E$ and $A$) used to simulate the spectra in figure \ref{fig:spin_transitions_rotations} and the extracted values for them from EPR experiments in \cite{chan1971electron}.}
  \begin{ruledtabular}
  \begin{tabular}{cccc}
  Ion & Parameter & Simulations & \cite{chan1971electron} \\
  \colrule
  V$^{2+}$ & S & 3/2 & 3/2 \\
  & I & 7/2 & 7/2 \\
  & $g$ & $2.09 \pm 0.05$ & 1.96 \\
  & $E$ (GHz) & $(\pm)0.58 \pm 0.1$ & $(\pm)$0.348 \\
  & $D$ (GHz) & $(\mp)4.5 \pm 0.1$ & $(\mp)$4.80 \\
  & $A$ (GHz) & $0.27 \pm 0.01$ & $(\pm)$0.25 $\pm$ 0.005 \\
  \colrule
  V$^{3+}$ & S & 1 & \\
  & I & 7/2 & \\
  & $g$ & $2.09 \pm 0.05$ & \\
  & $E$ (GHz) & $(\pm)1.2 \pm 0.1$ & \\
  & $D$ (GHz) & $(\pm)8.0 \pm 0.1$ & \\
  & $A$ (GHz) & $0.27 \pm 0.01$ & \\
  \colrule
  X & S & 3/2 & \\
  & I & 0 & \\
  & $g$ & $2.02 \pm 0.04$ & \\
  & $E$ (GHz) & $(\pm)0.7 \pm 0.1$ & \\
  & $D$ (GHz) & $|D| < 7.2$ & \\
  & $A$ (GHz) & NA & \\
  \end{tabular}
  \end{ruledtabular}
\end{table}

\begin{figure*}[t]
  \centering
  \includegraphics[width=\textwidth]{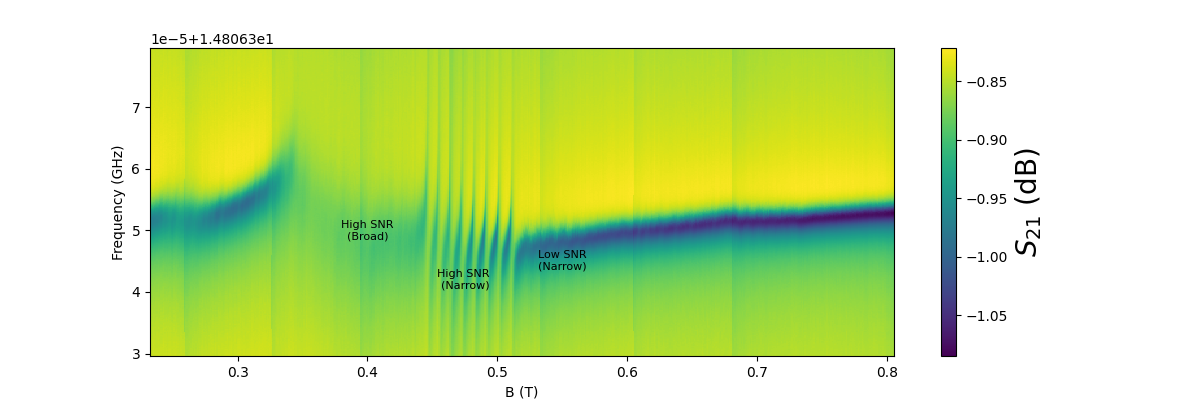}
  \caption{A colour plot of the microwave transmission power through the cavity as a function of input frequency and external DC magnetic field amplitude. The plot show examples of the 3 types of points plotted in figure \ref{fig:spin_transitions_rotations} with 3 different symbols: high SNR broad interactions (red squares), high SNR narrow interactions (green circles) and low SNR narrow interactions (blue circles).}
  \label{fig:cplot_interactions}
\end{figure*}

Interaction points arising from unidentified impurities were also observed in the experiment. In Figure~\ref{fig:spin_transitions_rotations}, these are plotted as blue squares. The blue solid lines in Figure~\ref{fig:spin_transitions_rotations} represent the fitted spectrum obtained by applying the model in equation~(\ref{eq:red_square_model}) to the blue square data points. This spectrum is labeled with ``X'' to indicate that the impurity type is unknown. Here, $S = 3/2$ and $I = 0$. The best-fit parameter values determined from this model are summarized in the last section of Table~\ref{tab:parameter_values}.

\begin{align} \label{eq:red_square_model}
  \hat{H} &= \mu_{B} g_{e} B_{z} \hat{S}_{z} + h D \left( S_{z}^{2} - \frac{1}{3}S(S+1) \right) \\
  &+ h E(S_{x}^{2} - S_{y}^{2}) \nonumber
\end{align}

Note that the parameter $D$ influences the splitting of the straight-line feature in the model but does not affect the curved feature until it exceeds a certain threshold. As a result, $D$ is reported as an upper bound in Table \ref{tab:parameter_values} for "X". An EPR color spectrum of these impurity-related points, along with the V$^{2+}$-related interaction points, is shown in Figure \ref{fig:cplot_interactions}.

\subsection{Coupling Strength and Spin Q Factor}

To characterize the hybridization of spin ensembles and photons in microwave spectroscopy experiments, one can assume that the Q factor of the spin ensemble dominates the Q factor of the resonator. This assumption holds when the coupling between resonator photons and spin ensembles is smaller than the inhomogeneous spin linewidth \cite{carvalho2017low}. Under this assumption, a two-mode magnetically coupled LCR oscillator model (labeled with $\omega_{1}(B)$ and $\omega_{2}(B)$ for each mode) can be developed \cite{tobar1991generalized, tobar1993effects}. The normal modes $\omega_{-}(B)$ and $\omega_{+}(B)$, which arise from the interaction of these two modes, can be derived by solving the following characteristic equation \cite{carvalho2017low}:

\begin{align} \label{eq:lcr2mode_normal_modes}
  \omega^{4} - \omega^2 (\omega_{1}^{2} + \omega_{2}^{2}) + \omega_{1}^{2} \omega_{2}^{2} - \Delta_{12}^{2} \omega_{1}^{2} \omega_{2}^{2} = 0
\end{align}

\begin{figure}[h]
  \centering
  \begin{subfigure}{0.44\textwidth}
      \centering
      \includegraphics[width=\linewidth]{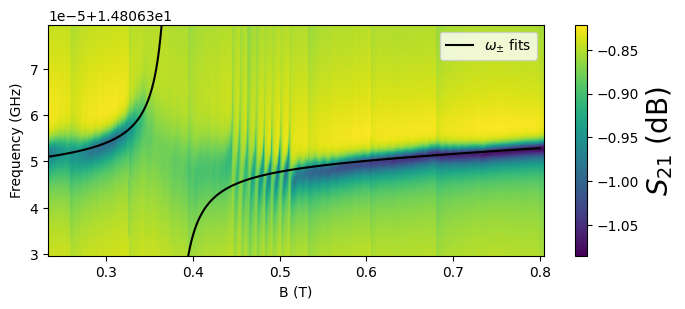}
      \caption{$\omega_{+}(B)$ and $\omega_{-}(B)$ fitted to a broad interaction point (a red square from figure \ref{fig:spin_transitions_rotations})}
      \label{fig:coupling_strength_L}
  \end{subfigure}
  \hfill
  \begin{subfigure}{0.44\textwidth}
      \centering
      \includegraphics[width=\linewidth]{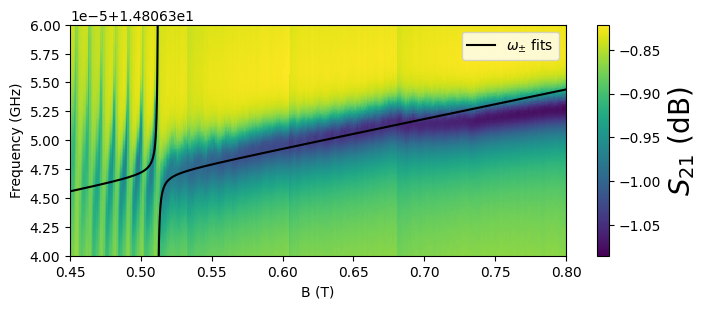}
      \caption{$\omega_{+}(B)$ and $\omega_{-}(B)$ fitted to a narrow interaction point (a green circle from figure \ref{fig:spin_transitions_rotations})}
      \label{fig:coupling_strength_s}
  \end{subfigure}
  \caption{$\omega_{+}(B)$ and $\omega_{-}(B)$ fittings for various types of points in figure \ref{fig:spin_transitions_rotations}}
  \label{fig:coupling_strengths}
\end{figure}

The values of $\omega_{+}(B)$ and $\omega_{-}(B)$ can then be fitted to experimental data to extract the coupling strength, given by $g_{\text{CS}} = \omega_{\text{WGM}} / 2\pi \times \Delta_{12}$. The extracted coupling strength values for the high SNR narrow and broad interactions in Figure \ref{fig:spin_transitions_rotations} are listed in Table \ref{tab:coupling_strengths}, with the corresponding fits shown in Figures \ref{fig:coupling_strength_L} and \ref{fig:coupling_strength_s}. Curve fitting could not be performed on the low SNR data points (see Figure \ref{fig:cplot_interactions}).

\begin{table}[h]
  \caption{\label{tab:coupling_strengths} Coupling strengths $g_{CS}$ for high SNR broad and narrow interactions from figure \ref{fig:spin_transitions_rotations}.}
  \begin{ruledtabular}
  \begin{tabular}{ccc}
  Fig. \ref{fig:spin_transitions_rotations} Interaction Label &
  $g_{CS}$ (kHz) \\
  \colrule
  High SNR Broad & $239.2 \pm 0.7$ \\
  High SNR Narrow & $31 \pm 2 $ \\
  \end{tabular}
  \end{ruledtabular}
\end{table}

The following expression is used to estimate the concentration of defect spins $n$ from the coupling strength $g_{CS}$, assuming uniform spin--mode coupling and a single dominant impurity species \cite{PhysRevB.84.060501}:

\begin{align} \label{eq:concentration_est}
    g_{CS} = g \mu_{B} \sqrt{\frac{\mu_{0} \omega_{0} n \xi_{\perp}}{4 \hbar}}
\end{align}

Here, $\mu_{B}$ is the Bohr magneton, $\mu_{0}$ is the permeability of free space, $\omega_{0}$ is the angular frequency of the microwave mode, $g$ is the effective Land\'e $g$-factor, and $\xi_{\perp}$ is the perpendicular magnetic filling factor ($\xi_{\perp} \approx 1$ for TM-WGMs) \cite{Bourhill:2016oqj}. Since the high SNR narrow interactions are attributed to V$^{2+}$ spin impurities, the corresponding spin concentration is estimated from equation~(\ref{eq:concentration_est}) to be $n = (1.0 \pm 0.2) \times 10^{10}$~cm$^{-3}$. To assess how this compares to the lowest concentration detectable by the current setup, following parameter values are assumed: $g_{CS} = 5$~Hz, $\omega_{0} = 2\pi \times 20$~GHz, and $\xi = 1$, which yields an approximate minimum detectable concentration of $n_{\text{min}} \approx 10^{7}$~cm$^{-3}$.

To solve for the Q factors of the normal modes ($Q_{-}(B)$ and $Q_{+}(B)$), a method similar to the one presented in \cite{tobar1993effects} for an n-mode system of mechanical oscillators can be used. This approach yields the following system of linear equations:

\begin{align} \label{eq:lcr2mode_Qpm}
  \begin{bmatrix} \frac{\omega_{+}}{Q_{+}} \\ \frac{\omega_{-}}{Q_{-}} \end{bmatrix} = \begin{bmatrix} 1 & 1 \\ \omega_{-}^{2} & \omega_{+}^{2} \end{bmatrix}^{-1} \begin{bmatrix} 1 & 1 \\ \omega_{2}^{2} & \omega_{1}^{2} \end{bmatrix} \begin{bmatrix} \frac{\omega_{1}}{Q_{1}} \\ \frac{\omega_{2}}{Q_{2}} \end{bmatrix}
\end{align}

\begin{figure}[h]
  \centering
  \includegraphics[width=0.46\textwidth]{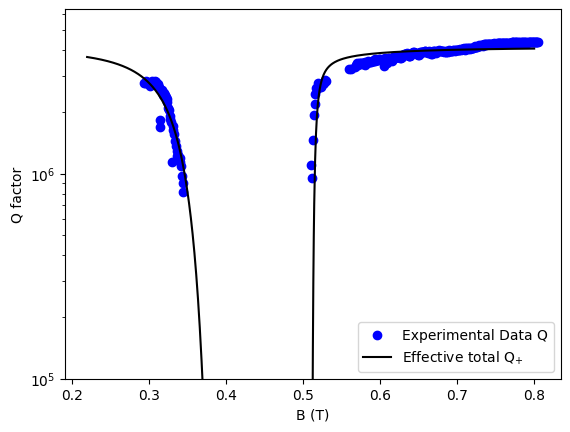}
  \caption{Fitted functions $Q_{+}(B, Q_{\text{WGM}}, Q_{\text{spin}})$ and $Q_{-}(B, Q_{\text{WGM}}, Q_{\text{spin}})$ overlaid on the experimental Q factors extracted from the mode interactions shown in Figs. \ref{fig:coupling_strength_L} and \ref{fig:coupling_strength_s}. The optimal parameter values obtained from the fit are listed in Table \ref{tab:q_factors}.}
  \label{fig:QvsBfit}
\end{figure}

The functions $Q_{+}(B, Q_{1}, Q_{2})$ and $Q_{-}(B, Q_{1}, Q_{2})$ can then be obtained from Equation (\ref{eq:lcr2mode_Qpm}). By fitting these functions to the experimentally measured Q factors extracted from the spectra, one can determine the Q factors of the WGM ($Q_{\text{WGM}}$) and the spin impurity ensemble ($Q_{\text{spin}}$). Figure \ref{fig:QvsBfit} shows the best-fit curves for $Q_{+}$ and $Q_{-}$ as a function of the magnetic field. The corresponding fitted oscillator Q values are listed in Table \ref{tab:q_factors}.

\begin{table}[h]
  \caption{\label{tab:q_factors} Best-fit values of the quality factor ($Q$) and linewidth ($\Gamma$) that most closely reproduce the experimental results extracted from the mode interactions shown in Figs. \ref{fig:coupling_strength_L} and \ref{fig:coupling_strength_s}.}
  \begin{ruledtabular}
  \begin{tabular}{ccc}
  Resonator &
  Q & 
  $\Gamma$ (kHz) \\
  \colrule
  Spin & $1.56 \times 10^{4} \pm 0.63 \times 10^{4}$ & $950 \pm 38$ \\
  WGM & $4.40 \times 10^{6} \pm 0.01 \times 10^{6}$ & $6.3 \pm 0.3$ \\
  \end{tabular}
  \end{ruledtabular}
\end{table}

\section{Conclusion}

Cryogenic microwave cavity spectroscopy was performed on a high-purity LiF crystal to investigate impurity populations. Characterizing these impurities is important for both engineering and fundamental physics applications; however, prior work on low-noise impurity characterization in LiF is limited. Vanadium impurities are known to be introduced during certain manufacturing processes of pure LiF \cite{anbinder1973spectrographic}, and previous EPR spectroscopy has shown that V$^{2+}$ and V$^{3+}$ ions with orthorhombic distortions can be present \cite{chan1970electron}. The EPR transitions of spin systems corresponding to these ions were simulated using an orthorhombic Hamiltonian, and the parameters $g$, $E$, $D$, and $A$ were fitted to the interaction data points. The simulations closely matched the experimental results, strongly supporting the presence of vanadium impurities in the sample. Due to the limited literature on other spin-$7/2$ impurities, such as cobalt, in pure LiF, the presence of such species in this sample cannot be confirmed or excluded based on the current data.

\section*{Author Declarations}

This work was funded by the ARC Centre of Excellence for
Engineered Quantum Systems, CE170100009, and Dark Matter
Particle Physics, CE200100008.

\subsection*{Conflict of Interest}

The authors have no conflicts to disclose.

\subsection*{Author Contributions}

\textbf{Steven Samuels}: Writing - original draft (lead); Writing - review \& editing (equal). \textbf{William Campbell}: Conceptualization (equal); Writing - review \& editing (equal). \textbf{Michael E. Tobar}: Conceptualization (equal); Writing - review \& editing (equal). \textbf{Maxim Goryachev}: Conceptualization (equal); Writing - review \& editing (equal).

\subsection*{Data Availability Statement}

The data that support the findings of this study are available from the corresponding author upon reasonable request.

\bibliography{aipsamp}

\end{document}